\documentclass[11pt]{article}
\usepackage{moriond,epsfig}
\def\be{\begin{equation}}
\def\ee{\end{equation}}
\def\bea{\begin{eqnarray}}
\def\eea{\end{eqnarray}}

\begin{document}
\vspace*{4cm}
\title{HIGH $Q^2$ STRUCTURE FUNCTIONS AT HERA}

\author{K. KORCSAK-GORZO\\ (On behalf of the H1 and ZEUS collaborations)}

\address{University of Oxford,\\Department of Particle Physics, Denys Wilkinson Building,\\Keble Road, Oxford OX1 3RH, England}

\maketitle

\abstracts{The latest results of H1 and ZEUS on the electroweak cross
  sections using polarised lepton beams are presented with the
  improvements in the parton density functions that result from them.
  The asymmetry of the neutral current cross sections between $e^- p$
  and $e^+p$ was measured and shows parity violation in agreement with
  the Standard Model. The dependence of the charged current cross
  section on the beam polarisation has been obtained and its
  extrapolation to full polarisation implies the absence of
  right-handed charged currents.  Constraints on the vector and axial
  vector couplings of the $u$ and $d$ quarks to the $Z$ boson from a
  combined QCD and electroweak fit are shown and the results from the
  combination of H1 and ZEUS data are reviewed.}

\section{Introduction}
The experiments H1 and ZEUS at the electron-proton collider HERA at
DESY collected data during the period 1992-2007 and have probed the
structure of the proton over a large kinematic region. The kinematic
scope of these experiments can be described in terms of the
four-momentum transfer squared, $Q^2$, between the colliding lepton
and the proton and the momentum carried by the struck quark,
Bjorken-x, and spans $0<Q^2<10^5$~GeV$^2$ and $10^{-6}<x<1$.

In the HERA-I running phase the collider provided unpolarised beams
until the luminosity upgrade in 2000, which marked the beginning of
the HERA-II running phase in which the lepton beam was longitudinally
polarised around the two experiments. H1 and ZEUS measured the cross
sections for neutral (NC) and charged current (CC) processes, from
which parton density functions (PDFs) were determined in global fits
at next-to-leading order in QCD. For an understanding of the structure
of the proton a precise measurement of the PDFs is crucial. With the
beginning of beam operation at the Large Hadron Collider (LHC)
imminent, the results from HERA are particularly important as the
constraints on the PDFs from this data are relevant in a significant
part of the LHC kinematic region.

\begin{figure}[h]
\begin{center}
\psfig{figure=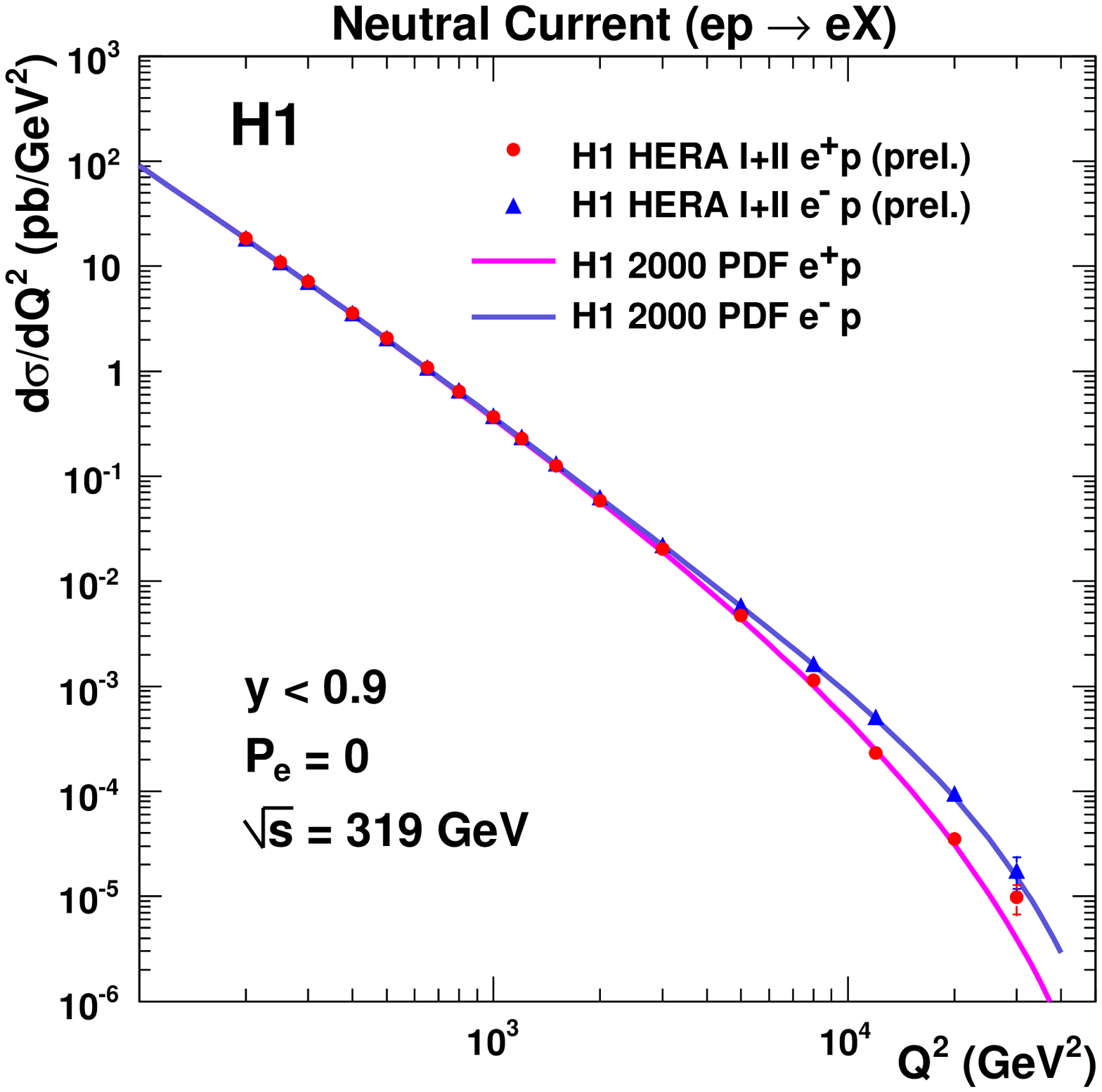,height=5.0cm}
\hspace{0.2cm}
\psfig{figure=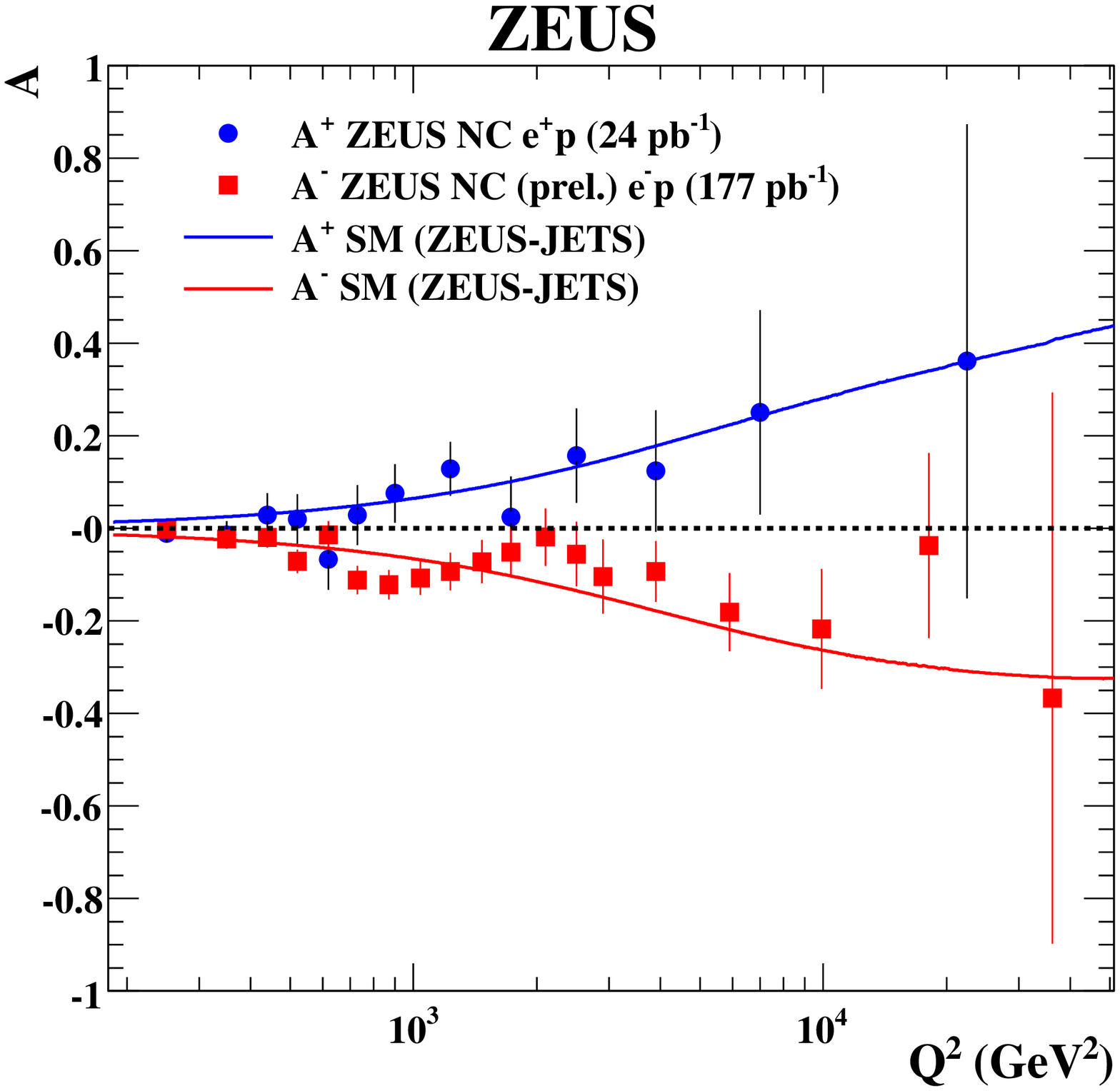,height=4.9cm}
\hspace{0.2cm}
\psfig{figure=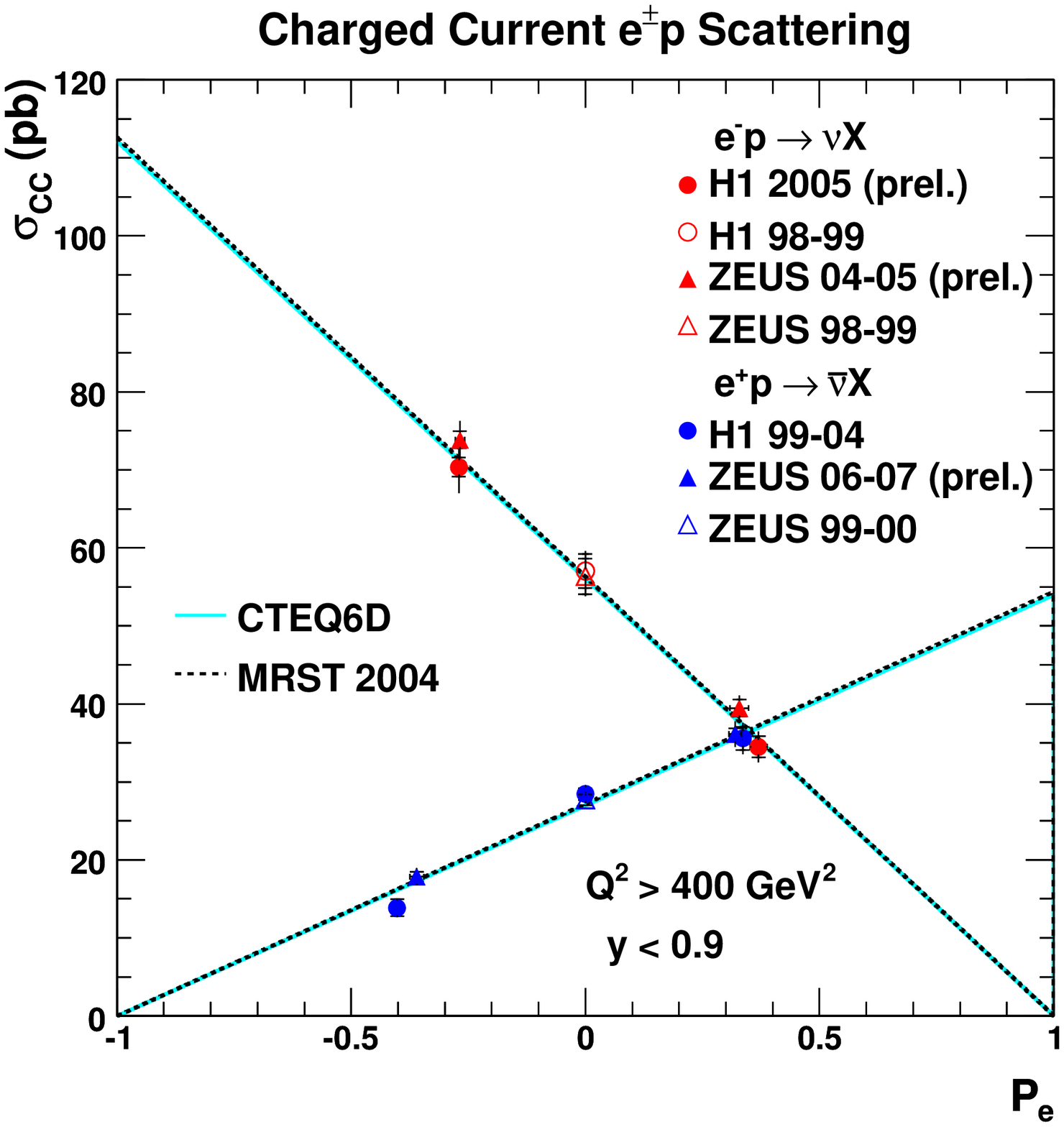,height=5.0cm}\\
(a)\hspace{5cm} (b)\hspace{5cm}(c)
\caption{(a) NC single differential cross sections versus $Q^2$ as
  measured by H1. (b) NC asymmetry in $e^+p$ and $e^-p$ measured by
  ZEUS. (c) Polarisation dependence of the CC cross sections measured
  by H1 and ZEUS.\label{fig:CCvsPOL}}
\end{center}
\end{figure}

\section{Electroweak cross sections and structure functions}

The cross section for the NC process is given by \cite{rced:xsec:dis}
\begin{equation}
\tilde{\sigma}^{e^\pm p} = \frac{xQ^4}{2\pi\alpha^2 Y_+}\frac{d^2\sigma(e^\pm p)}{dxdQ^2}=F_2^\pm(x,Q^2)\mp \frac{Y_-}{Y_+}xF_3^\pm(x,Q^2)-\frac{y^2}{Y_+}F_L^\pm(x,Q^2),
\end{equation}
where $\alpha$ is the fine structure constant, $Y_\pm = 1\pm(1-y)^2$
and $y$ can be obtained from the expression $Q^2=sxy$ using the centre
of mass energy, $\sqrt{s}$. The above equation shows the relation
between the cross section and the structure functions of the proton.
These are $F_2$, which dominates the cross section and is related to
the sum of the quark and antiquark distributions, $xF_3$, which
becomes significant at high $Q^2$ and is proportional to the
difference of the quark and antiquark distributions and $F_L$, the
longitudinal structure function, which contributes at high $y$ and
gives a measure of the gluon density in the proton. However, in the
kinematic range relevant here $F_L$ is small and can be neglected.

H1 results on the NC single differential cross sections $d\sigma/dQ^2$
using 270~pb$^{-1}$ $e^+p$ and 165~pb$^{-1}$ $e^-p$ data from the
HERA-I and II running periods are shown in Fig.~\ref{fig:CCvsPOL}a.
The measurements were obtained in the kinematic regime of
$200<Q^2<30\,000$~GeV$^2$ and $y<0.9$. At high $Q^2$ the $e^-p$ cross
sections are larger than the $e^+p$ cross sections as expected from
the positive contribution of $xF_3$ in the former case. H1 used these
cross sections to perform a form factor analysis from which an upper
limit on the quark radius was obtained of $0.74 \cdot 10^{-18}$~m at
95\% CL \cite{H1.FormFactor}.

The NC cross section contains a dependence on the lepton polarisation,
$P_e$, which enters through the structure functions.
This polarisation dependence can be expressed in terms of the
asymmetry, $A$, which is defined at full polarisation as
\begin{equation}
A = \frac{\sigma(P_e=+1)-\sigma(P_e=-1)}{\sigma(P_e=+1)+\sigma(P_e=-1)}.
\end{equation}
The asymmetry as measured by ZEUS using 24~pb$^{-1}$ of $e^+p$ and
177~pb$^{-1}$ of $e^-p$ data is presented in Fig.~\ref{fig:CCvsPOL}b
and compared to the SM predictions using the ZEUS-JETS PDFs
\cite{ZEUSJETS}. The divergence between the asymmetry of $e^+p$ and
$e^-p$ at high $Q^2$ is a clear indication of parity violation in the
NC processes.

The leading order CC cross section, neglecting weak radiative
corrections can be expressed as:
\begin{equation}
\frac{d^2\sigma_{CC}(e^{\pm}p)}{dxdQ^2} = (1\pm P_e) \frac{G_F^2}{2\pi}(\frac{M_W^2}{M_W^2 + Q^2})^2 [u' + c' + (1-y)^2(d' + s' + b')],\\
\end{equation}
where $G_F$ is the Fermi coupling constant and $u',c',d',s'$ represent
the quark densities of the respective flavour and stand for
$u,c,\bar{d},\bar{s},\bar{b}$, in the case of $e^-p$ scattering and
for $\bar{u},\bar{c},d,s,b$ in the case of $e^+p$.  These expressions
reveal that the dominant PDFs at high $x$, $u$ and $d$, are best
probed by deep inelastic scattering in $e^-p$ and $e^+p$,
respectively. The expected linear dependence of the CC cross section
on the polarisation has been measured by H1 and ZEUS and can be seen
in Fig.\ref{fig:CCvsPOL}c. The extrapolation of the cross sections to
full polarisation indicates the absence of right-handed CC processes.
The upper limit on the mass of a right-handed $W$ boson was obtained
by H1 to be 186~GeV/$c^2$ at 95\% CL \cite{H1.RH.W.limit}.

\begin{figure}[h]
\begin{center}
\hspace{0.5cm}
\psfig{figure=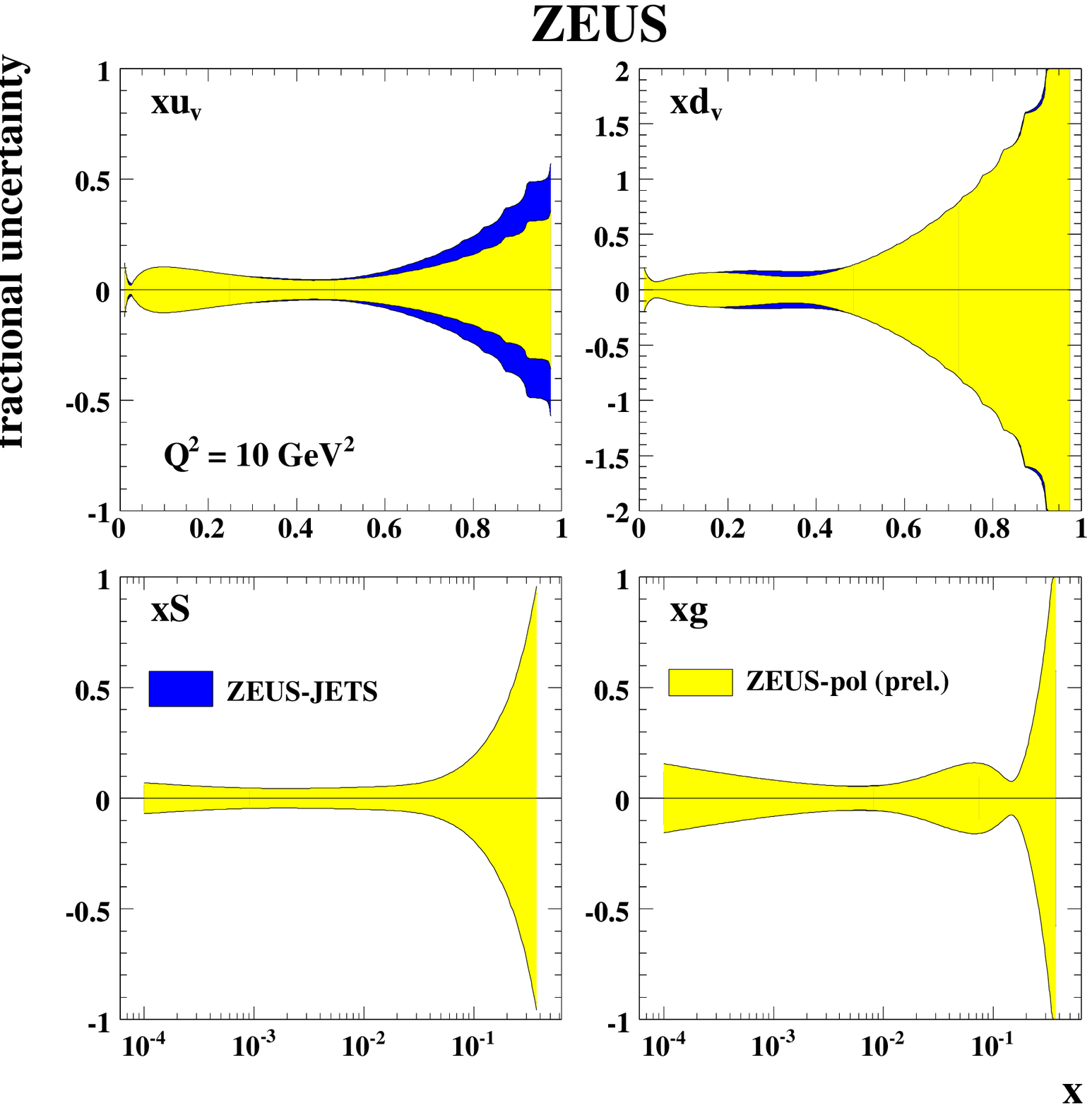,height=7.2cm}
\hspace{5mm}
\psfig{figure=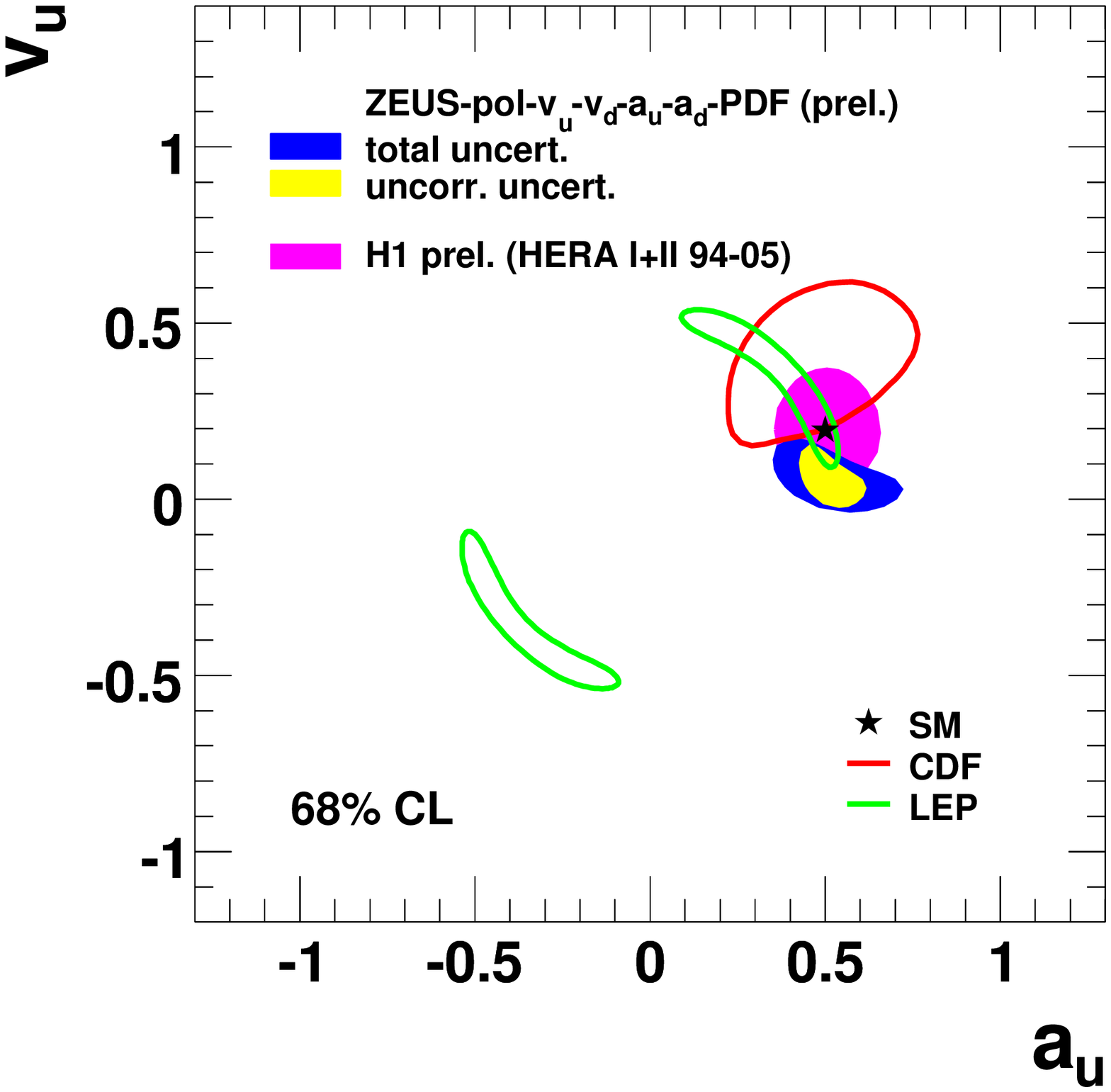,height=7.2cm}\\
(a)\hspace{7cm}(b)
\caption{ (a) Fractional uncertainties of the up-valence,
  down-valence, sea and gluon PDFs using ZEUS-JETS and
  ZEUS-pol.\label{fig:PDFs}, (b) Axial and vector couplings of the
  $u$-quark obtained from a combined QCD and electroweak
  fit.\label{fig:coupl}}
\end{center}
\end{figure}

\section{Electroweak and QCD fits using polarised data}
Utilising the measurements of the electroweak cross sections H1 and
ZEUS obtained PDFs as well as quark couplings from a simultaneous QCD
and electroweak fit. In addition to the data sets used in previous
fits the polarised data from HERA-II was also taken into account.
Figure~\ref{fig:PDFs}a compares the fractional uncertainties on
individual PDFs between the old ZEUS-JETS based on HERA-I ZEUS data
only and the new ZEUS-pol PDFs \cite{ZEUS.pol} which also uses HERA-II
ZEUS $e^-p$ data. The $u$ valence-quark PDF benefits the most from the
addition of the new $e^-p$ data, which leads to a significant
improvement of the fractional uncertainties in the high $x$ region.

Limits on the axial ($a$) and vector ($v$) couplings of the $u$ and
$d$ quark were obtained from a fit in which $a_u$, $a_d$, $v_u$, $v_d$
were all treated as free parameters. The results for the couplings of
the $u$-quark are shown in Fig.~\ref{fig:coupl}b. Compared to previous
H1 and ZEUS constraints the axial couplings improved under the
addition of an order of magnitude more $e^-p$ data, whereas the vector
couplings benefited the most from the polarisation of that new data
set. The limits set by H1 and ZEUS are competitive with limits
obtained by CDF and LEP.

\section{Combination of H1 and ZEUS data}
\begin{figure}[h]
\begin{center}
\psfig{figure=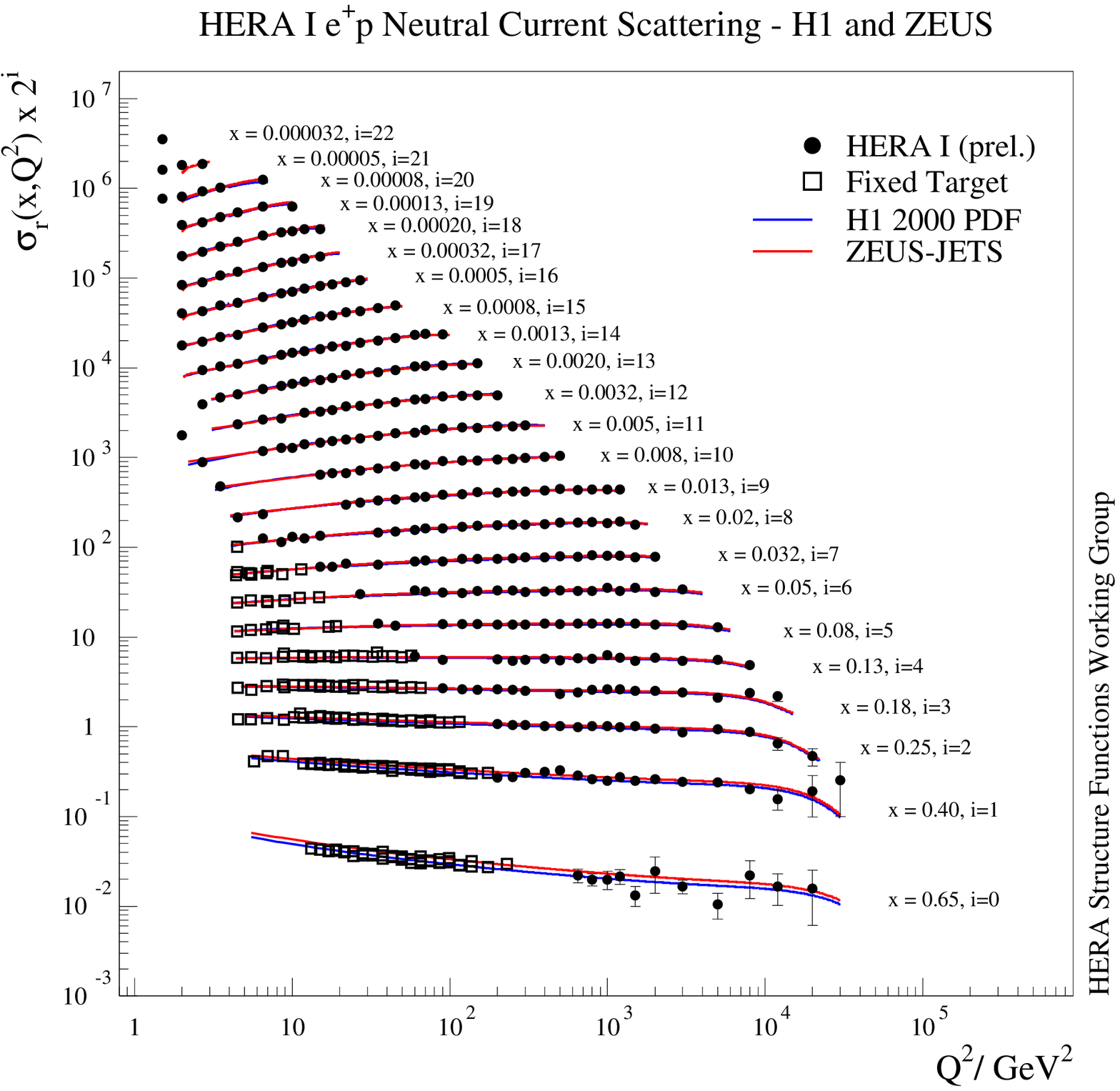,height=7.9cm}
\psfig{figure=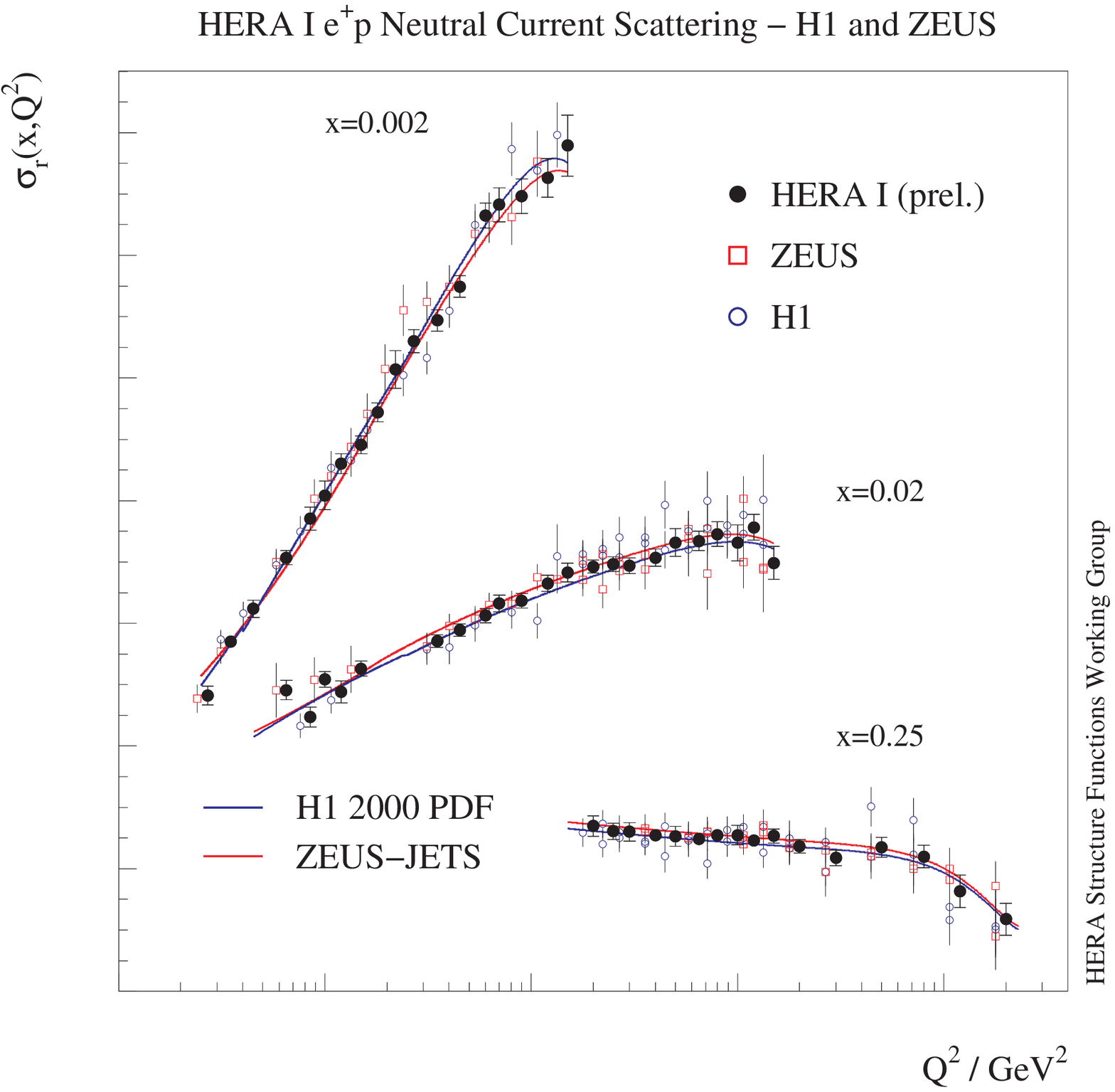,height=7.9cm}\\
(a)\hspace{7cm}(b)
\caption{(a) H1 and ZEUS combined NC cross sections compared to
  results from fixed target experiments and SM predictions using the
  ZEUS-JETS and H1 2000 PDFs. (b) Enlarged plot for x=0.002, 0.02,
  0.25 highlighting the reduced uncertainty on the combined results
  (black) compared to individual H1 (blue) and ZEUS (red)
  results.\label{fig:H1.ZEUS.combi}}
\end{center}
\end{figure}

To exploit the full potential of HERA the results of H1 and ZEUS have
to be combined in a model independent way. The published HERA-I deep
inelastic cross sections from these experiments have thus been merged
for $Q^2>1.5$~GeV$^2$.  The method used is described in detail
elsewhere~\cite{glazov}. The combination procedure, in which
systematic correlations are taken into account, leads to a significant
reduction of the overall uncertainty.  The results for the combined H1
and ZEUS NC cross sections are shown in Fig.~\ref{fig:H1.ZEUS.combi}.
The strength of the method lies in the reduction of the uncertainty
both in the region of low $Q^2$, where systematic uncertainties
dominate, as well as at high $Q^2$, where the combined measurement
benefits from improved statistics. Systematic uncertainties are
reduced by allowing the correlated uncertainties to vary coherently in
the averaging procedure. The uncertainties are largely uncorrelated
between H1 and ZEUS, thus, a cross calibration between the experiments
is achieved resulting in an overall reduction of the systematic
errors. This is highlighted in Fig.~\ref{fig:H1.ZEUS.combi}b, which
shows an enlarged version of the plot on the left. The overall
uncertainty on the combined data points labelled 'HERA I', shown in
black, is smaller than the uncertainty on individual H1 and ZEUS data
points across the whole range of $Q^2$~ \cite{H1.ZEUS.combi}.

Further work needs to be done to combine the HERA-II data from H1 and
ZEUS. But upon completion of that project the 'final word' from HERA
will comprise 1~fb$^{-1}$ and will significantly contribute to the
precise knowledge of the proton PDFs for the upcoming experiments at
the Large Hadron Collider.

\end{document}